\documentstyle[12pt, fleqn]{article}
\begin{document}

\title{ \begin{flushright}
        {\normalsize 
         DOE/ER/40561--248--INT96--00--117 \\ }
        \end{flushright}
        \vspace{0.7 cm}
        \bf Relativistic Calculations for Photonuclear
	   Reactions (III): A Consistent Relativistic
	   Analysis of the $\left(e, e^\prime p\right)$ and
	   $\left(\gamma, p\right)$ Reactions\thanks{
           Work supported in part by the Natural Sciences
           and Engineering Research Council of Canada} }

\author{ {\bf J.I. Johansson} \\
              Institute for Nuclear Theory, University of Washington \\
              Box 351550, Seattle, WA, 98195-1550 \\
              and \\
              Department of Physics, University of Manitoba \\
              Winnipeg, Manitoba, R3T 2N2 \\
              and \\
         {\bf H.S. Sherif } \\
              Department of Physics, University of Alberta \\
              Edmonton,  Alberta, Canada T6G 2J1 \\
              and \\
         {\bf G.M. Lotz } \\
              Augustana University-College \\
              Camrose, Alberta, Canada T4V 2R3}

\date{\today}

\maketitle

\newpage

\begin{abstract}

Relativistic calculations for the quasifree electron scattering
process $\left(e, e^{\prime}p\right)$ and the direct knockout
contribution to $\left(\gamma, p\right)$ reactions are presented.
The spectroscopic factors determined from the former reaction
are used to fix the magnitude of the knockout contribution to  
the $\left(\gamma, p\right)$ reaction at 60 MeV.
The results obtained for several nuclei indicate that the 
knockout contributions are much larger in magnitude and hence 
closer to the data than predicted in an earlier comparison based
on non-relativistic calculations.
We discuss the sensitivity of the results to the choice of
parameters for the binding and final state interactions.
We find these uncertainties to be more pronounced at the larger
missing momenta explored by the $\left(\gamma, p\right)$ reaction.
The implications of  the present results for the size of contributions
due to meson exchange currents are discussed.
 \end{abstract}


\newpage

\section{Introduction}      \label{intro}
Reactions initiated by electromagnetic probes play a central 
role in our understanding of the structure-dynamics of 
nucleons and nuclei. In the latter case both quasifree
$\left(e, e^{\prime}p\right)$ and $\left(\gamma, p\right)$ 
reactions have been extensively studied. These two 
processes have a great deal in common in that both reactions
result in the removal of a single proton from the target,
via interaction with a photon, exciting essentially the same
residual states.
The first of the reactions is mediated by virtual photons, whereas
the second is initiated by real ones. 
Moreover the two reactions complement each other in providing
information about different momentum regions of single particle
wave functions. 

It has long been recognized that the similarity between the 
two reactions could be very useful in enhancing our 
understanding of the reaction mechanisms\cite{FO77,BO84}.
A recent study by Ireland and van der Steenhoven \cite{IS94} 
has compared results of non-relativistic DWIA calculations 
for $\left(e, e^{\prime}p\right)$ and $\left(\gamma, p\right)$ 
reactions on a number of light and medium-weight nuclei. 
Under the reasonable assumption that the mechanism for the former 
reaction is well understood, calculations were performed first
for this reaction and used in conjunction with the available
data to determine the relevant spectroscopic factors and bound
state parameters.
The resulting parameters were then used to constrain similar 
calculations for the $\left(\gamma, p\right)$ reaction for a 
photon energy of 60 MeV.
The objective was to quantify the importance of the direct
knockout (DKO) mechanism to this process and hence to
assess the extent of the contributions due 
to meson exchange current effects.
The study revealed the surprising result that the constrained 
DKO calculations fell considerably below the data in most cases.
The authors report an average value of 5.8 for the ratio between
data and calculations.

The above results led Ireland and van der Steenhoven
to explore possible
contributions from meson exchange currents to the
$\left(\gamma, p\right)$ reaction in a simple model.
Using Seigert's theorem they estimated these contributions
in the plane wave limit where they could determine a factor
given by the
ratio of the full cross section to the DKO cross section.
When this factor was applied to the distorted wave
DKO calculations, the agreement with data showed considerable 
improvement.
The authors then concluded that meson exchange current
contributions to $\left(\gamma, p\right)$ reactions, in this
energy region, must be significant.

The purpose of the present paper is to report on similar
comparative calculations carried out within a relativistic 
mean field approach. We have recently carried out 
relativistic distorted wave calculations for the DKO
contributions to $\left(\gamma, p\right)$ reactions
\cite{LS88}.
There have also been several similar calculations for the
$\left(e, e^{\prime}p\right)$  reaction
\cite{Mc90,Ud93,JO92,HJS95}.
Moreover, recent studies by Hedayati-Poor {\it et al.}
\cite{HJS95,HS94,HJS95_ii}
as well as other authors \cite{Ud93,Ud95,JO94}
have concentrated on the question of the differences between
relativistic and non-relativistic calculations for these two
closely related reactions.
In particular, the investigations reported in 
\cite{HJS95,HS94,HJS95_ii} have pointed to the existence
of subtle medium modifications to the interaction hamiltonians
in the relativistic approach, which are absent in the
corresponding non-relativistic treatment.
This, in addition to the fact that the non-relativistic calculations
referred to above require such large meson exchange contributions,
suggests that a reanalysis in the relativistic framework is advisable.
With the strong medium effects alluded to above, it is
possible that the role of meson exchange currents could be strongly
modified.

We carry out a comparison between the two reactions mentioned above
along similar lines to those used by Ireland and van der Steenhoven.
Relativistic calculations are carried out for
$\left(e, e^{\prime}p\right)$  reactions on a number of nuclei.
The spectroscopic factors determined from comparison with the data 
are used to make predictions for
$\left(\gamma, p\right)$ reactions on the same nuclei.
These predictions are compared directly to the cross section data.
 
We outline the relativistic calculations for quasifree
electron scattering and the direct knockout contribution
to the $\left(\gamma, p\right)$ reaction in section \ref{rel_obs}.
We then provide discussion of the specific ingredients of the
models and results of the calculations for the two reactions
in section \ref{disc}.
Our conclusions are given in section \ref{concl}.

\section{The Relativistic Calculations}   \label{rel_obs}

The relativistic calculation of the amplitude in the one photon
exchange model for the $\left(e, e^{\prime}p\right)$ process
is outlined in reference \cite{HJS95}. 
The main results are given briefly here with some change of
notation in order to highlight the similarities between the
two reactions considered here.
We do not include the Coulomb distortion in the leptonic part
of the amplitude.
This will only be important for heavy nuclei \cite{Mc90,GP87}
which will not be considered in the present paper.

The relativistic expression for the differential cross
section leading to a specific final state of the residual
nucleus can be written as
\begin{eqnarray}
 \frac{d^3\sigma}{d\Omega_{p} d\Omega_f dE_f}  &=& 
   \frac{2} {\left( 2 \pi \right)^{3}}  \frac{\alpha^{2}} {\hbar c}
   { \left[ \frac{ \left( m_e c^{2} \right)^{2}
                   M c^{2} \;
                   {\left| \mbox{\boldmath{$p$}}_{p} \right|} c}
                 { \left( q c \right)^{4} }
             \frac{ {\left| \mbox{\boldmath{$p$}}_{f} \right|} c}
                  { E_i }
            \right] }
    \frac{ c } {v_{rel} } \frac{1}{R}
  \nonumber
   \\  & & \times { \frac{ {\cal S}_{J_i J_f} (J_B) }{ 2J_B + 1 } }
    \sum_{ \mu M_B \nu_{f} \nu_{i} }
    { \left| e_{\nu_{f} \nu_{i}}^{\alpha}
             N_{\alpha}^{\mu M_B} \right| }^2 ,
  \label{cross}
\end{eqnarray}
where $\nu_i$ and $\nu_f$ are the spin projections of the incoming
and outgoing electrons respectively, while $M_B$ and $\mu$ are the
spin projections of the bound and continuum protons.
The 4-momenta of the initial and final electrons are
$p_{i}$ and $p_{f}$ respectively, while the final proton
4-momentum is $p_{p}$. The 4-momentum of the exchanged photon
is $q$ and is calculated as the difference between the
initial and final electron 4-momenta $q = p_{i} - p_{f}$.
The 4-momentum of the recoil nucleus is $p_{R}$ and the initial
4-momentum of the struck proton is denoted $p_{m}$, which is 
often called the missing momentum.
The recoil factor $R$, was not included in reference \cite{HJS95}
but we do include it here for completeness.
$R$ is given in any frame by \cite {FM84}
\begin{eqnarray}
   R = 1 - \frac{ E_{p} } { E_{R} }
           \frac{1} {\left| \mbox{\boldmath{$p$}}_{p} \right|^2}
           \mbox{\boldmath{$p$}}_{p} \cdot
           \mbox{\boldmath{$p$}}_{R}     .
   \label{recoil_fact}
\end{eqnarray}
The function $N_{\alpha}^{\mu M_B}$ is
\begin{eqnarray}
  N_{\alpha}^{\mu M_B} = \int d^3x \; 
        \Psi_{\mu}^\dagger \left( p_{p}, \mbox{\boldmath{$x$}} \right)
        \Gamma_{\alpha}
        \Psi_{J_{B}, M_{B}} \left( \mbox{\boldmath{$x$}} \right)
        \exp \left( i \mbox{\boldmath{$q$}} \cdot
                      \mbox{\boldmath{$x$}} \right)     .
   \label{nuc_mat}
\end{eqnarray}
where the wave functions of the continuum and bound nucleons,
denoted $\Psi_{\mu}$ and $\Psi_{J_{B}, M_{B}}$ respectively,
are solutions of the Dirac equation containing appropriate
potentials \cite{LS88}.
The $4 \times 4$ matrix $\Gamma_{\alpha}$, operating on the
nucleon spinors is given in Eq. (2.8) of reference
\cite{HJS95} and the four-vector which comes from the electron
vertex $e_{\nu_{f} \nu_{i}}^{\alpha}$, is given in Eq.
(2.9) of that same reference.


The distorted momentum distribution (referred to as `reduced
cross section' by Ireland and van der Steenhoven in reference
\cite{IS94}) is obtained from the cross section
given above through division by a kinematic factor and the cross
section for the elementary $e+p \rightarrow e+p$ process.
We write \cite{TDFJ,BGP}:
\begin{eqnarray}
  \rho \left( \mbox{\boldmath{$p$}}_{m} \right)
    = \frac{ \frac{ d^3\sigma}
                  {  d\Omega_{p} d\Omega_f dE_f}  }
           {  E_p \left| \mbox{\boldmath{$p$}}_{p} \right|
               d\sigma_{ep}^{cc1}  }    .
   \label{mom_dist}
\end{eqnarray}
In the figures and the following discussion we refer to this
simply as the momentum distribution.
The space-like portion of the missing 4-momentum,
in the impulse approximation, is the negative of the recoil
nucleus 3-momentum in the lab frame
$\mbox{\boldmath{$p$}}_{m} = - \mbox{\boldmath{$p$}}_{R}$.
The free cross section, $d\sigma_{ep}^{cc1}$,
is that which de Forest denotes $cc1$
\cite{TDFJ} and is evaluated using the kinematics of the
quasifree process with his prescription for the 
energy of the bound nucleon,
$p_{m}^{0}
    = \left[ \left| \mbox{\boldmath{$p$}}_{m} \right|^{2}
      + M^{2} \right]^{1/2}$.
This prescription is used only in calculating the free cross
section $cc1$ which divides our calculated cross section in Eq. 
(\ref{mom_dist}) since $cc1$ divides the experimentally measured
cross sections to obtain the momentum distributions which
are presented as experimental data,
see for example \cite{St88, Le94}.
For the distorted wave calculations of the cross section
given in Eq. (\ref{cross}) we use a slightly
different prescription for the missing energy, namely,
$p_{m}^{0}
    = \left[ \left| \mbox{\boldmath{$p$}}_{m} \right|^{2}
      + \left( M - E_{s} \right)^{2} \right]^{1/2}$,
where $E_{s}$ is the energy required to remove the bound
nucleon from the target nucleus.
Thus the bound proton is off-shell due to the modification
of its mass by the separation energy.


The cross section for the $\left(\gamma, p\right)$ reaction
in which the residual nucleus is left in a definite final
state can be written as \cite{LS88}
\begin{eqnarray}
  \frac{d\sigma}{d \Omega_{p}} 
     &=& \frac{ \alpha }{ 4 \pi }  \frac{ M c^2 }{ \hbar c }
         \frac{ {\left| \mbox{\boldmath{$p$}}_{p} \right|} c }
              {E_{\gamma} } 
         \frac{ c } {v_{rel} }
         \frac{1}{R}
         { \frac{ {\cal S}_{J_i J_f} (J_B) }{ 2J_B + 1 } }
         \sum_{\mu M_B r }
         { \left| \epsilon_{r}^{\alpha}
                  N^{\mu M_B}_{\alpha} \right| }^2
    \label{gp_cross}
\end{eqnarray}
where for the direct knockout contribution, the function
$N^{\mu M_B}_{\alpha}$ is the nuclear matrix element of
Eq. (\ref{nuc_mat}).
The 4-vector $\epsilon_{r}^{\alpha}$ is the photon
polarization vector with two polarization states $r$.
The recoil factor $R$, has exactly 
the same form as that of Eq. (\ref{recoil_fact}) but the
kinematics are now those of the $\left(\gamma, p\right)$
reaction.

In the comparison with experimental data it is not necessary
to define a common function for both the
$\left(e, e^{\prime}p\right)$ and $\left(\gamma, p\right)$
reactions.
Ireland and van der Steenhoven presented their results in terms
of common reduced cross sections in reference \cite{IS94}.
The appeal of this procedure stems from the fact that the
non-relativistic amplitudes for the two reactions reduce,
in the plane wave limit, to simpler forms that depend on the
momentum wave function of the struck nucleon.
The relativistic amplitudes, however, do not lend themselves
to a similar simplification.
The important point in this regard, is that the two reactions
probe
complementary momentum regions of the nuclear wave function:
the $\left(e, e^{\prime}p\right)$ reaction for low momenta,
and the $\left(\gamma, p\right)$ reaction for higher momenta.
In that sense it is extremely interesting to carry out a joint 
analysis of the two processes.
This however does not restrict the way in which the individual
data for each reaction is presented.
Thus in the next section we compare our calculations for
$\left(e, e^{\prime}p\right)$ with data for the momentum
distribution because that is the way the data are usually 
presented. Note that we are simply dividing our 
calculated differential cross section by the same factor
by which the experimental differential cross section has been
divided so the representation is irrelevant.
We then compare the calculated and experimental 
cross sections for the $\left(\gamma, p\right)$ reaction
and check whether the predicted relativistic results
are in better agreement than those obtained
non-relativistically.

\section{Discussion}         \label{disc}

We begin our discussion with the elements common to calculations
of the two reactions.
The bound state protons are described by solutions of a Dirac
equation containing the relativistic Hartree potentials of
Horowitz and Serot \cite{HS86}.
These potentials, obtained through self-consistent calculations,
usually underbind the protons in the states we consider by
1-3 $MeV$.
The $rms$ radii, however, are in reasonable agreement with those
found by Ireland and van der Steenhoven \cite{IS94} as well as
values obtained by other authors \cite{St88,Le94,Kr89,He88}.

The continuum proton in the final state is described by 
solutions of a Dirac equation containing complex 
phenomenological optical potentials obtained from fits to
proton elastic scattering data \cite{COPE}.
There are several sets of potentials available, some of
which are energy dependent (E-dep) and constructed from a fit
to data for a specific nucleus, such as $^{12}C$, $^{16}O$
and $^{40}Ca$, in the proton kinetic energy
range of $\sim$ 25 $MeV$ to 1 $GeV$. 
Other potentials are parameterized in terms of target mass as
well as proton energy (E+A-dep) and can be used to generate
potentials for which no proton elastic scattering data exist.
We shall perform calculations using both types of potentials.

\begin {table} 
\begin {center}
\begin{tabular}  { || c | c || c | c || c || }   \hline \hline
  nucleus   &   nucleon   
            & $S^{R}$       & $S^{R}$
            &  $S^{nR}$                     \\
            &   state     
            &  E-dep      &  E+A-dep
            &                               \\    \hline \hline
  $^{12}C$  & $1p_{3/2}$  
            &      2.00     &     2.24
            & $1.825^{\cite{IS94}}$
                                            \\    \hline
  $^{16}O$  & $1p_{1/2}$  
            &      1.26     &     1.38
            & $1.124^{\cite{IS94}}$
                                            \\    \cline{2-5}
            & $1p_{3/2}$  
            &      2.24     &     2.60
            & $\sim 2.2^{\cite{Le94}}$
                                            \\    \hline
 $^{40}Ca$  & $1d_{3/2}$  
            &      3.24     &     3.12
            & $2.698^{\cite{IS94}}$
                                            \\    \hline
  $^{51}V$  & $1f_{7/2}$  
            &      n/a     &      0.46
            & $0.384^{\cite{IS94}}$
                                            \\    \hline \hline
\end{tabular}
\end {center}
\caption{Spectroscopic factors extracted from
         $\left(e, e^{\prime}p\right)$ data. The 
         superscripts $R$ and $nR$ on the spectroscopic factors
         refer to relativistic and non-relativistic calculations
         respectively.}
\end {table}

Given the potentials discussed above, the only parameters 
left to determine are the spectroscopic factors.
In order to obtain spectroscopic factors $S^{R}$, the relativistic
calculations for quasifree electron scattering are normalized
by eye to the right-hand peak of the experimental data, where the 
error bars tend to be smallest.
Results of our calculations for $\left(e, e^{\prime}p\right)$
reactions are plotted along with the experimental data in parts
(a) of Figs. 1 to 5.
In all cases the shape of the data is well described by our
calculations indicating that the Hartree wave functions are
providing a reasonable description of the bound proton
in the range of missing momenta considered.
The spectroscopic factors obtained are recorded in table 1.

Column 3 of table 1 shows the spectroscopic factors extracted
from calculations in which the final proton distortion is provided
by E dependent potentials of Cooper {\it et al.} \cite{COPE}
for a particular target nucleus.
Column 4 of Table 1 shows the spectroscopic factors extracted
from calculations in which the final proton distortion is provided
by {\it `fit 1'} of the E and A dependent potentials of that
same reference.
The last column of table 1 shows the spectroscopic factors
obtained through the non-relativistic analysis of Ireland and
van der Steenhoven \cite{IS94}.
The non-relativistic spectroscopic factor for knockout of the 
$1p_{3/2}$ proton from $^{16}O$ is obtained from the analysis
of Leuschner {\it et al.} \cite{Le94}
(the data for this state were not considered in reference
\cite{IS94}).
These spectroscopic factors are in close agreement with others
obtained through similar non-relativistic analyses
\cite{St88,Le94,Kr89,He88}.
Note that the relativistic spectroscopic factors are slightly
larger than those obtained non-relativistically, in agreement
with other analyses (see for example Udias
{\it et al.} \cite{Ud93} and references therein).
One important feature of the present analysis is the difference
in spectroscopic factors obtained from calculations involving
the different relativistic potentials.
This is a measure of our uncertainty in the knowledge of the 
continuum proton wave function.
We see differences of up to fifteen percent between the
spectroscopic factors obtained using the E dependent
potentials from those obtained using the E and A
dependent potentials.
We find similar differences when we use different bound
state potentials with the E dependent optical potentials.
Even though there are differences in the extracted
spectroscopic factors arising from sensitivity to the 
input ingredients of the model, the resulting fits to the
$\left(e, e^{\prime}p\right)$ data are all of similar
quality.
These uncertainties should be kept in mind when drawing
conclusions concerning the magnitude of the DKO contribution
to $\left(\gamma, p\right)$ reactions.
 
Relativistic calculations of the direct knockout contribution
to $\left(\gamma, p\right)$ reactions are carried out using the
same relativistic Hartree potentials and distorting potentials
used in the electron scattering calculations, as well as the
same relativistic spectroscopic factors (see Table 1).
We stress that there are no parameters left to adjust when
comparing the results of the $\left(\gamma, p\right)$
calculations to the data.
The resulting cross sections can be regarded as relativistic 
predictions for the knockout contribution to the
$\left(\gamma, p\right)$ reaction.
The results are shown in parts (b) of Figs. 1-5.
The solid curves in all figures show the results
of calculations for the reactions using the 
Hartree binding potentials and {\it `fit 1'} of the
E and A dependent optical potentials discussed earlier.
The dashed curves in Figs. 2 and 3 use E dependent optical
potentials along with Hartree binding, while the dotted curves
result from utilizing a Woods-Saxon binding potential.
The curves of Fig. 5 result from using three
different E and A dependent potentials.

In order to emphasize the complementary momentum regions
explored by the two reactions considered we have included an
insert in Fig. 1(b) showing the missing momentum as a function
of the proton lab angle for the $\left(\gamma, p\right)$
reaction with a $^{12}C$ target.
Note that the the lowest missing momentum available in the
$\left(\gamma, p\right)$ reaction is near the maximum of
the missing momenta for which data are shown
for the $\left(e, e^{\prime} p\right)$ reaction.
In moving to larger nuclei with the same incident photon
energy in the lab the missing momentum simply scales 
upward slightly.
  
In comparing our relativistic calculations with the 
$\left(\gamma, p\right)$ data it is striking that our
calculations lie slightly above the data in two cases
(for the $^{12}C$ and $^{16}O$ targets shown in Figs.
1 and 3), and below in two cases (for the $^{16}O$ and
$^{40}Ca$ targets shown in Figs. 2 and 4).
In the fifth case in which the target is $^{51}V$ the
calculations lie within the error bars of the data.
This is contrary to the results of the non-relativistic 
analyses of Ireland and van der Steenhoven who found that
DKO calculations for $\left(\gamma, p\right)$ reactions
lie consistently below the data; they reported an average
factor of 5.8 for light nuclei \cite{IS94}.
The largest differences between data and calculations occur
for the lightest targets in both the non-relativistic and
relativistic analyses. However the results of our relativistic
calculations lie much closer to the data than the
corresponding non-relativistic results for these light nuclei.

At this point it may be useful to comment on some aspects of the
potential sensitivities in the present calculations.
The Hartree potentials
result in a binding energy that is slightly smaller than
the experimental value. Alternate calculations for $^{16}O$
were performed using Woods-Saxon binding potentials,
which reproduce the experimental binding energy and also 
provide an $rms$ radius for the bound state that is within one
percent of that found from the Hartree potentials.
We find little sensitivity to the results between the two
potentials for the $\left(e, e^{\prime}p\right)$ reaction
(see part (a) of Figs. 2 and 3).
This is because the momentum space wave functions for the bound
states are very similar in the low momentum region explored by
the $\left(e, e^{\prime}p\right)$ reaction and only begin to
show differences in the higher momentum region available via the
$\left(\gamma, p\right)$ reaction.
Figures 2(b) and 3(b) show that this is indeed the case, 
with a range of difference in the neighborhood of $30 \%$.
We have also done calculations in which the depth of the
Dirac-Hartree scalar potential is varied in order to reproduce
the experimental binding energy. This requires a change in depth
of the potential of less than $3 \%$ but can yield a change in
the $\left(\gamma, p\right)$ cross section of up to $30 \%$.
  
Previously we pointed out some sensitivity to changes in 
the global optical potentials leading to different values of
spectroscopic factors determined from
$\left(e, e^{\prime}p\right)$ data.
We have done calculations for the $^{12}C$ and $^{51}V$ 
targets using the three available E and A dependent fits 
\cite{COPE}, and results are shown for the $^{51}V$ target
in Fig. 5.
For the $\left(e, e^{\prime}p\right)$ reaction the results
change by less than five percent amongst the three potentials.
The $\left(\gamma, p\right)$ reaction shows slightly more
sensitivity particularly for angles larger than $90^{\circ}$,
where the momentum transfer is large.
However, there are no striking differences in shape or magnitude
due to changes in the optical potentials.
Similar results are obtained for $^{12}C$.

It is evident from the above discussion that
relativistic calculations for $\left(\gamma, p\right)$
reactions do show some sensitivity to changes in the
ingredients, and an estimate of these sensitivities
combined leads to a possible variation in the magnitude of the
cross sections by up to a factor of two, along with some
variation in shape.
It has also been noted by Harty {\it et al.} \cite{Ha95} that there
are unexplained inconsistencies within the set of experimental
$\left(\gamma, p\right)$ data leading to differences of factors
of up to two between differential cross sections at the same 
angle and photon energy while the systematic errors are
quoted as $10 \%$ to $22 \%$.
Within these uncertainties our relativistic calculations are
much closer to the $\left(\gamma, p\right)$ data than
the non-relativistic calculations of Ireland and
van der Steenhoven.

\section{Conclusions}                   \label{concl}

In this paper we have presented a comparative study of
$\left(e, e^{\prime}p\right)$ and $\left(\gamma, p\right)$
reactions for several nuclei.
The objective is to make a quantitative assessment of the 
contribution from the direct knockout mechanism to the
$\left(\gamma, p\right)$ reactions in the relativistic
approach and to compare this with the non-relativistic results
of Ireland and van der Steenhoven.
The $\left(e, e^{\prime}p\right)$ reaction was used to fix the 
spectroscopic factors which were subsequently used to make
predictions for the $\left(\gamma, p\right)$ reaction.

In the course of our study we have looked at the dependence
of the results, for both $\left(e, e^{\prime}p\right)$ and
$\left(\gamma, p\right)$ reactions, on the type of bound state
used and on the optical potentials describing the final state
interactions of the outgoing proton.
We find little sensitivity to the choice of binding potential
in the $\left(e, e^{\prime}p\right)$ reaction;
this is largely due to the low missing momenta covered by the
present experimental data.
The $\left(\gamma, p\right)$ reaction explores a higher
range of missing momenta, and we find slightly more sensitivity
to the choice of binding and optical potentials.
Although the optical potentials we use are the best currently
available there is clearly room for improvement in their precise
specification.
Note however that we find less variation amongst the potentials
used here, as measured by the spectroscopic factor
extracted via the $\left(e, e^{\prime}p\right)$ reaction,
than found in earlier calculations \cite{LS88,HJS95_ii}.

With the above comments in mind, our relativistic analysis shows
that the DKO contribution lies much closer to the data,
compared to the non-relativistic calculations of Ireland and
van der Steenhoven,
who have found that the DKO calculations lie consistently below
the data by an average factor of 5.8.
The factors required to bring our calculations to
the data are consistently less than 2.
It should be noted that our present conclusions are in agreement
with the findings of Ryckebusch et al. \cite{Ry92} 
(see also Bobeldijk {\it et al.} \cite{Bo95}) who have done
non-relativistic RPA calculations of meson exchange contributions
to $\left(\gamma, p\right)$ reactions.
They have found that meson exchange is not the dominant
contributor for the $\left(\gamma, p\right)$ reaction leading
to the ground state of the residual nucleus, while it can
modify the DKO cross section by up to a factor of two when the
residual nucleus is excited to some low lying states.

We conclude that the direct knockout contributions to the
$\left(\gamma, p\right)$ reaction are much closer to the data
than is indicated in the non-relativistic calculations of
Ireland and van der Steenhoven.
This implies that meson exchange currents may not play
as dominant a role in our relativistic calculations,
although they may certainly contribute
significantly in obtaining the correct shape and magnitude
of the cross section.
In order to fully understand the role of the various
competing mechanisms, it is desirable in this regard to
carry out complete relativistic calculations,
including meson exchange current corrections,
and to compare the results with the data over a wide range
of photon energies.

\section*{Acknowledgements}

We would like to thank L. Lapik\'{a}s and G. van der Steenhoven
for providing tables of the data used in the present paper.
We are grateful to M. Hedayatipoor for help in checking the 
numerical calculations and for useful discussions.
Two of us (JIJ and HSS) would like to thank the Institute for
Nuclear Theory at the University of Washington for their
hospitality during the completion of this work.

\newpage

\section* {Figure Captions}

\noindent FIG. 1. Knockout of a $1p_{\frac{3}{2}}$ proton 
from a $^{12}C$ target leading to the $^{11}B$
ground state.
Hartree bound state wave functions are used \cite{HS86} 
and the proton optical potentials are E+A-dep, fit 1, 
from reference \cite{COPE}.
(a) Momentum distribution for the reaction
$^{12}C\left(e, e^{\prime}p\right)^{11}B_{g.s.}$.
The energy of the incident electron is 481.1 $MeV$,
and the kinetic energy of the detected proton is fixed at
70 MeV with parallel kinematics.
The data are from reference \cite{St88}.
(b) Cross section for the reaction 
$^{12}C\left(\gamma, p\right)^{11}B_{g.s.}$.
The photon energy is 60 $MeV$.
The data are from reference \cite{Sp90}.

\vspace{2.5 mm}
\noindent FIG. 2. Knockout of a $1p_{\frac{1}{2}}$ proton 
from a $^{16}O$ target leading to the $^{15}N$
ground state.
(a) Momentum distribution for the reaction
$^{16}O\left(e, e^{\prime}p\right)^{15}N_{g.s.}$.
The energy of the incident electron is 456 $MeV$,
and the kinetic energy of the detected proton is fixed at
90 MeV with parallel kinematics.
The data are from reference \cite{Le94}.
(b) Cross section for the reaction 
$^{16}O\left(\gamma, p\right)^{15}N_{g.s.}$.
The photon energy is 60 $MeV$.
The data are from reference \cite{Mi95}.
Dashed curve --- Hartree binding potential and
                 E-dep optical potential for $^{16}O$.
Dotted curve --- Woods-Saxon binding potential and
                 E-dep optical potential for $^{16}O$.
Solid curve ---  Hartree binding potential and
                 E+A-dep optical potential, fit 1.

\vspace{2.5 mm}
\noindent FIG. 3. Same as Fig. 2 but for knockout of a
$1p_{\frac{3}{2}}$ proton leading to the
${\frac{3}{2}}^{-}$ excited state at 6.3 $MeV$ in $^{15}N$.

\vspace{2.5 mm}
\noindent FIG. 4. Knockout of a $1d_{\frac{3}{2}}$ proton 
from a $^{40}Ca$ target leading to the $^{39}K$
ground state.
Hartree bound state wave functions are used \cite{HS86} 
and the proton optical potentials are E+A-dep, fit 1, 
from reference \cite{COPE}.
(a) Momentum distribution for the reaction
$^{40}Ca\left(e, e^{\prime}p\right)^{39}K_{g.s.}$.
The energy of the incident electron is 460 $MeV$,
and the kinetic energy of the detected proton is fixed at
100 MeV with parallel kinematics.
The data are from reference \cite{Kr89}.
(b) Cross section for the reaction 
$^{40}Ca\left(\gamma, p\right)^{39}K_{g.s.}$.
The photon energy is 60 $MeV$.
The data are from reference \cite{Ab95}.

\vspace{2.5 mm}
\noindent FIG. 5. Knockout of a $1f_{\frac{7}{2}}$ proton 
from a $^{51}V$ target leading to the $^{50}Ti$
ground state.
Hartree bound state wave functions are used \cite{HS86} 
and the proton optical potentials are E+A-dep, 
from reference \cite{COPE}.
(a) Momentum distribution for the reaction
$^{51}V\left(e, e^{\prime}p\right)^{50}Ti_{g.s.}$.
The energy of the incident electron is 461 $MeV$,
and the kinetic energy of the detected proton is fixed at
70 MeV for $p_m < 140 MeV/c$ and
100 MeV for $p_m > 140 MeV/c$, with parallel kinematics.
The data are from reference \cite{He88}.
(b) Cross section for the reaction 
$^{51}V\left(\gamma, p\right)^{50}Ti_{g.s.}$.
The photon energy is 60 $MeV$.
The data are from reference \cite{Ir93}.
Solid curve --- fit 1.
Dashed curve --- fit 2.
Dotted curve --- fit 3.

\begin{figure}
\begin{picture}(1100,400)(0,0)
\includegraphics{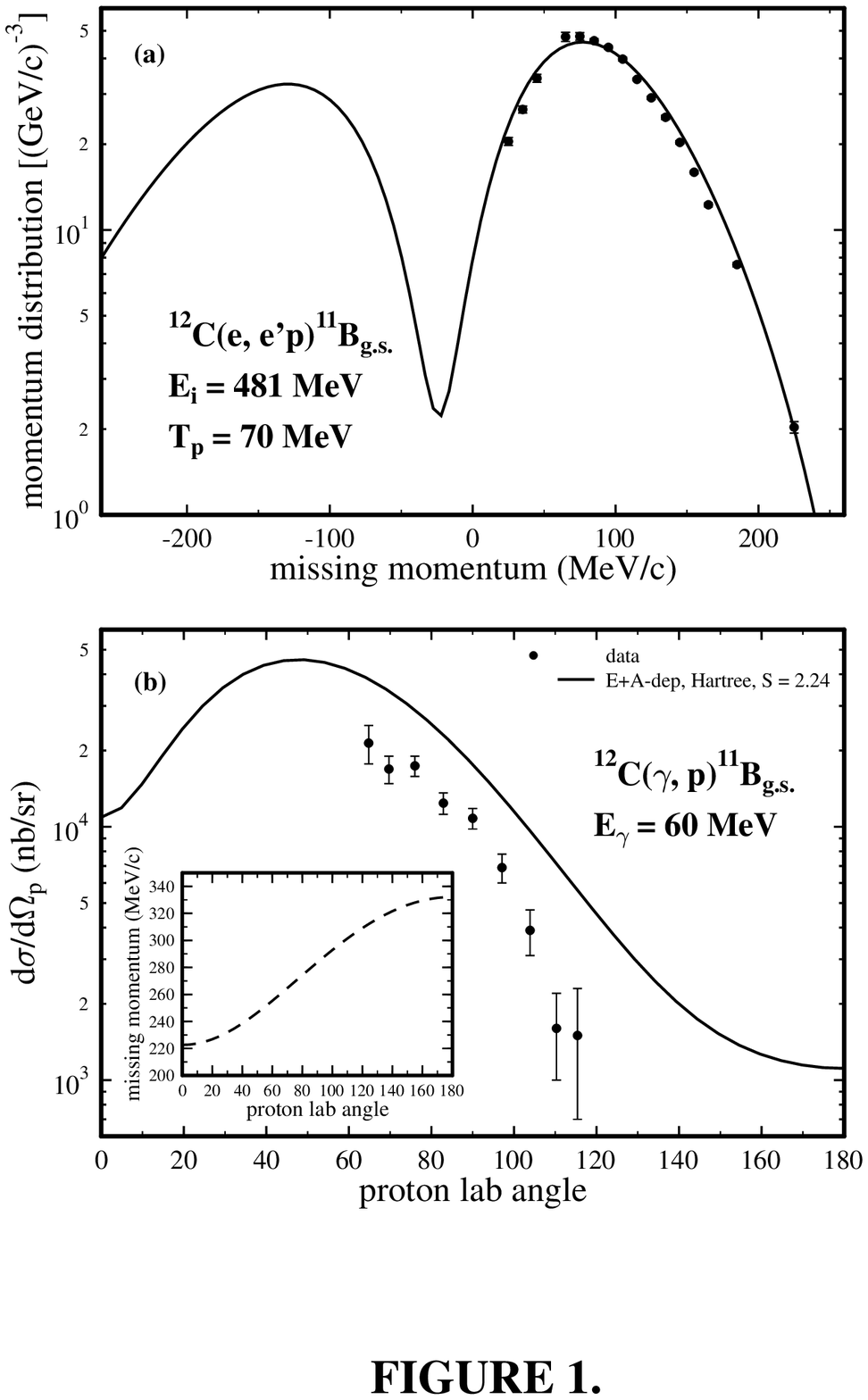}
\end{picture}
\end{figure}

\begin{figure}
\begin{picture}(1100,400)(0,0)
\includegraphics{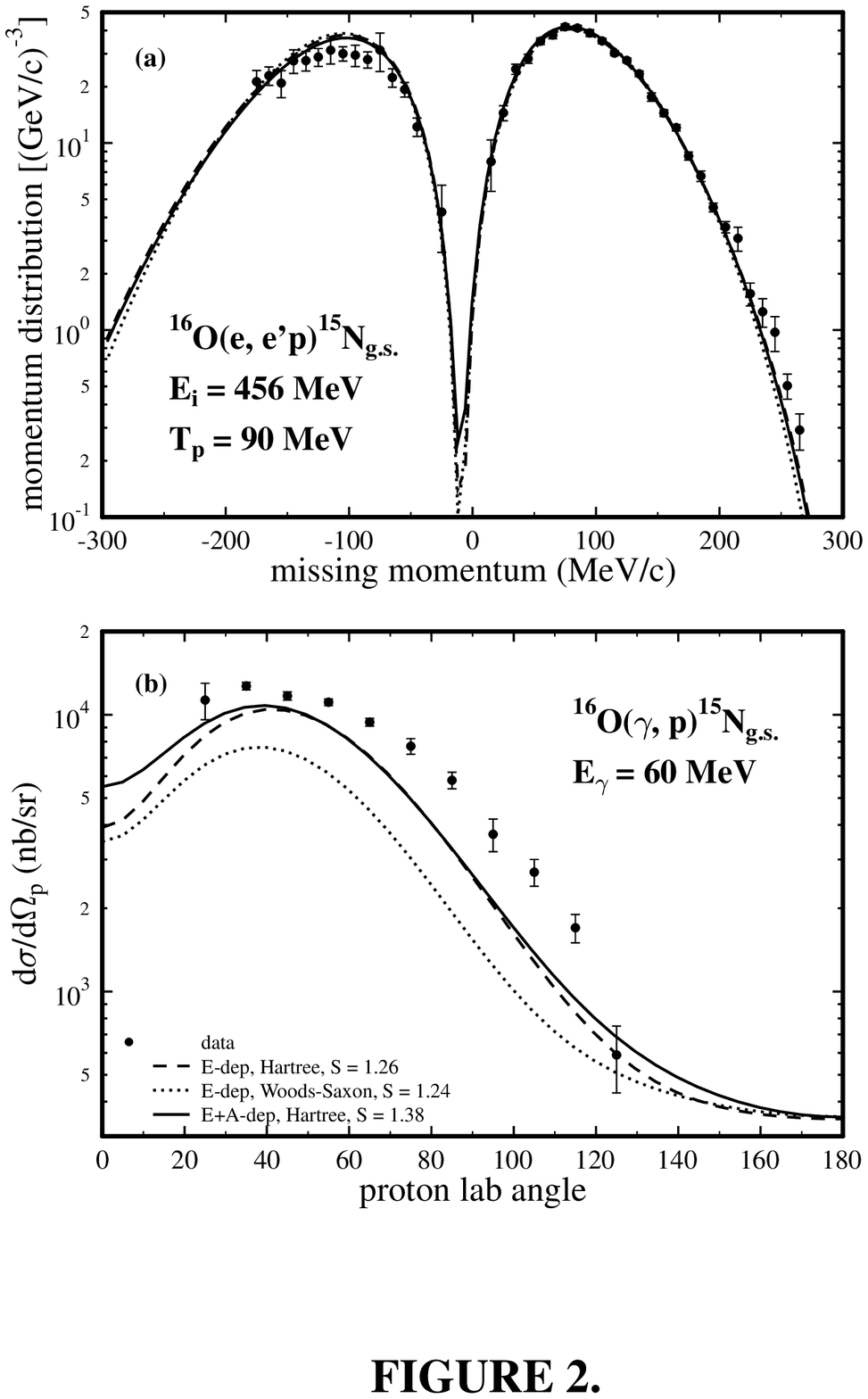}
\end{picture}
\end{figure}

\begin{figure}
\begin{picture}(1100,400)(0,0)
\includegraphics{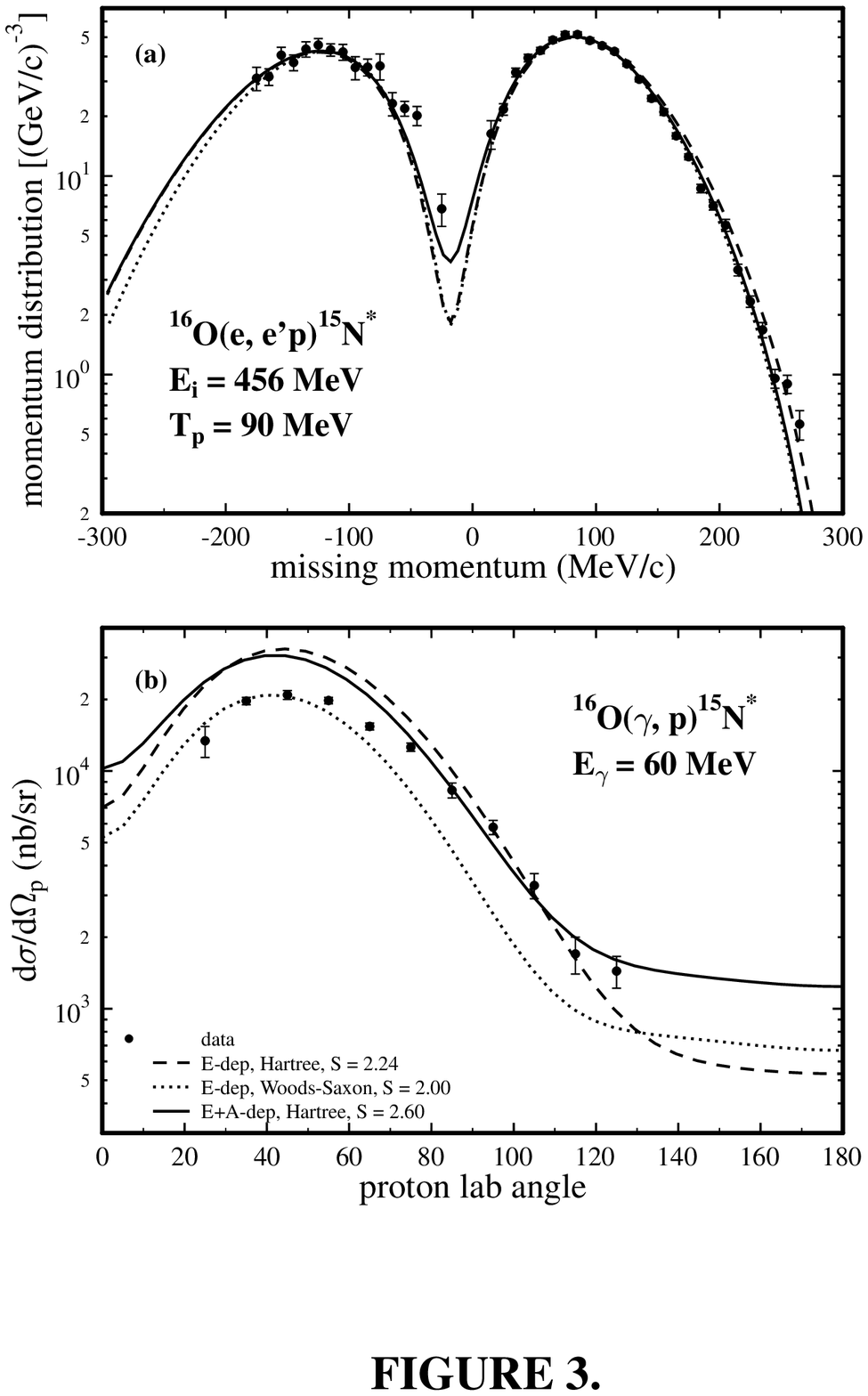}
\end{picture}
\end{figure}

\begin{figure}
\begin{picture}(1100,400)(0,0)
\includegraphics{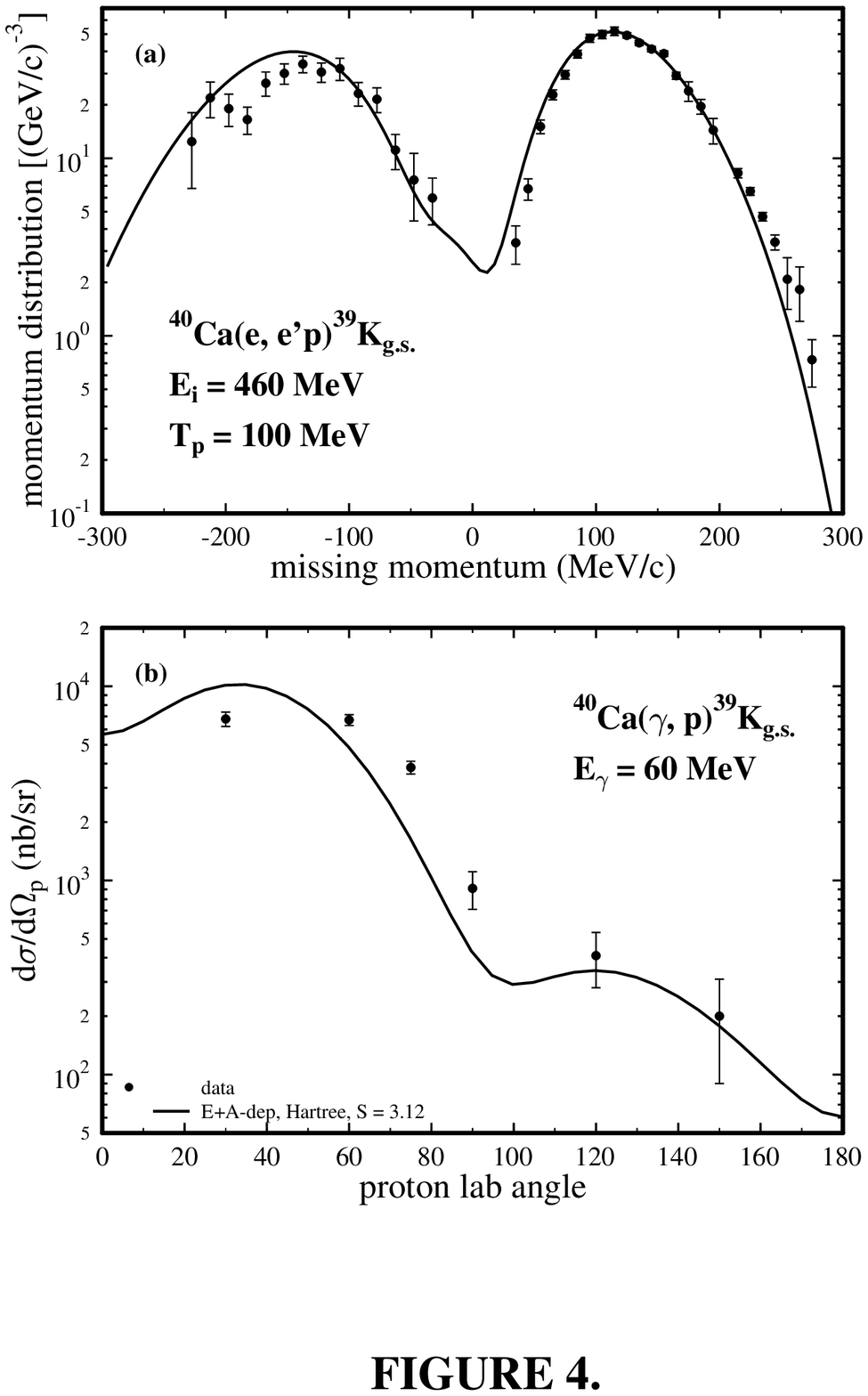}
\end{picture}
\end{figure}

\begin{figure}
\begin{picture}(1100,400)(0,0)
\includegraphics{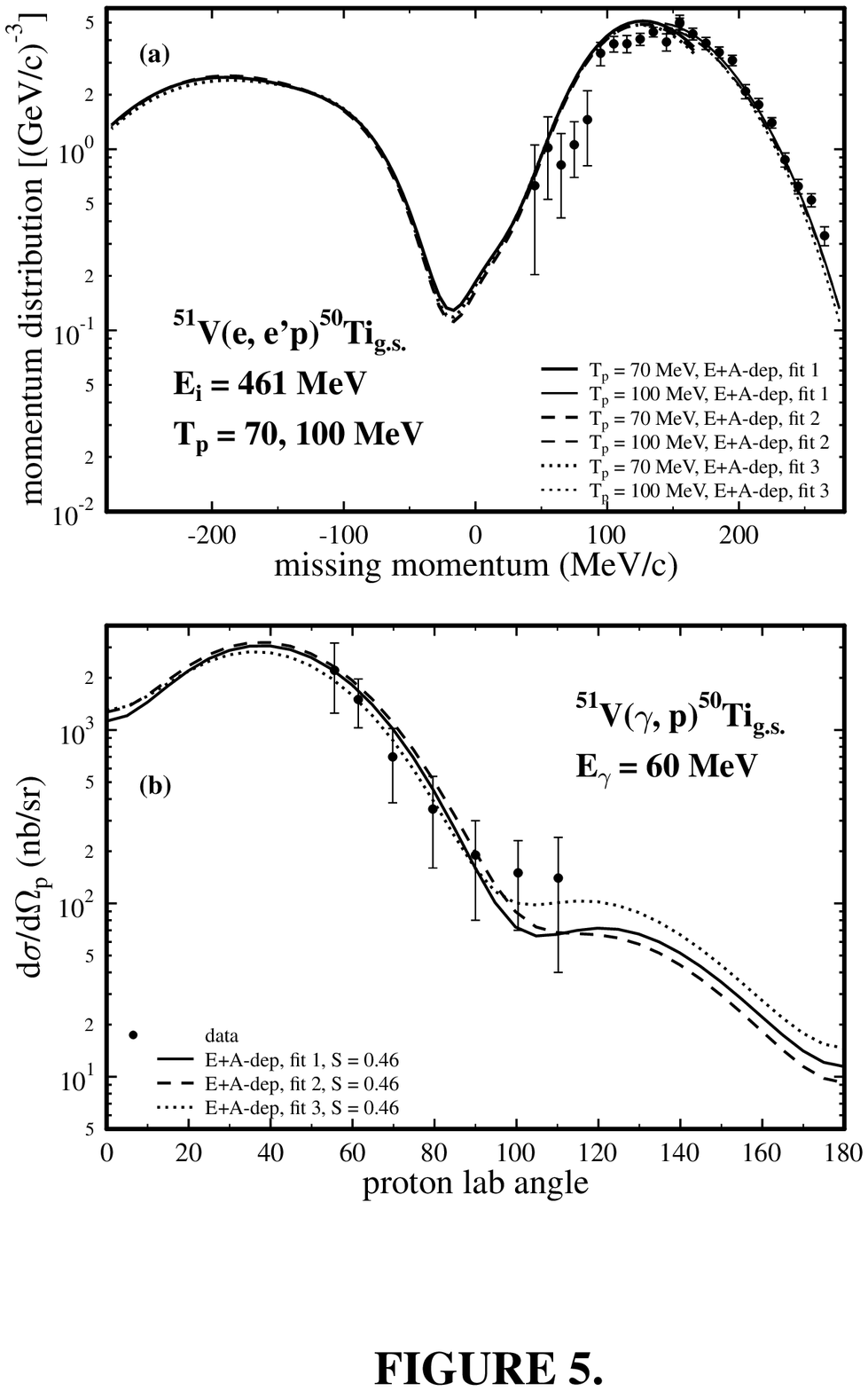}
\end{picture}
\end{figure}


\newpage

\begin {thebibliography} {99}
\bibitem {FO77} D.J.S. Findlay and R.O. Owens,
                Nucl. Phys. {\bf A292} (1977) 53.
\bibitem {BO84} S. Boffi, R. Cenni, C. Giusti and F.D. Pacati,
                Nucl. Phys. {\bf A420} (1984) 38.  
\bibitem {IS94} D.G. Ireland and G. van der Steenhoven,
                Phys. Rev. C {\bf  49} (1994) 2182.
\bibitem {LS88} G.M. Lotz and H.S. Sherif,
                Phys. Lett. B {\bf 210} (1988) 45; and
                Nucl. Phys. {\bf A537} (1992) 285.
\bibitem {Mc90} J.P. McDermott,
                Phys. Rev. Lett. {\bf 65} (1990) 1991.
\bibitem {Ud93} J.M. Udias, P. Sarriguren, E. Moya de Guerra,
                E. Garrido and J.A. Caballero,
                Phys. Rev. C {\bf 48} (1993) 2731.
\bibitem {JO92} Y. Jin, D.S. Onley and L.E. Wright,
                Phys. Rev. C {\bf 45} (1992) 1311.
\bibitem {HJS95} M. Hedayati-Poor, J.I. Johansson and H.S. Sherif,
                 Phys. Rev. C {\bf  51} (1995) 2044.
\bibitem {HS94} M. Hedayati-Poor and H.S. Sherif,
                Phys. Lett. B {\bf 326}  (1994) 9.
\bibitem {HJS95_ii} M. Hedayati-Poor, J.I. Johansson and H.S. Sherif,
                 Nuc. Phys. {\bf A593} (1995) 377.
\bibitem {Ud95} J.M. Udias, P. Sarriguren, E. Moya de Guerra,
                E. Garrido and J.A. Caballero,
                Phys. Rev. C {\bf 51} (1995) 3246.
\bibitem {JO94} Yanhe Jin and D.S. Onley,
                Phys. Rev. C {\bf 50} (1994) 377.
\bibitem {GP87} C. Giusti and F.D. Pacati,
                Nucl. Phys. {\bf A473} (1987) 717.
\bibitem {FM84} S. Frullani and J. Mougey,
                {\it Advances in Nuclear Physics}, edited by 
                J.W. Negele and E. Vogt, {\bf 14} (1984) 1.
\bibitem {TDFJ} T. de Forest Jr.,
                Nucl. Phys. {\bf A392} (1983) 232.
\bibitem {BGP}  S. Boffi, C. Giusti and F.D. Pacati,
                Nucl. Phys. {\bf A336} (1980) 416;
                Nucl. Phys. {\bf A336} (1980) 427.
\bibitem {St88} G. van der Steenhoven {\it et al.},
                Nuc. Phys. {\bf A480} (1988) 547.
\bibitem {Le94} M. Leuschner {\it et al.},
                Phys. Rev. C {\bf 49} (1994) 955.
\bibitem {HS86} C.G. Horowitz and B.D. Serot,
                Nucl. Phys. {\bf A368} (1986) 503.
\bibitem {Kr89} G.J. Kramer {\it et al.},
                Phys. Lett. B {\bf 227} (1989) 199.
\bibitem {He88} J.W.A. den Herder {\it et al.},
                Nuc. Phys. {\bf A490} (1988) 507.
\bibitem {COPE} E.D. Cooper, S. Hama, B.C. Clark and R.L. Mercer,
                Phys. Rev. C {\bf 47} (1993) 297.
\bibitem {Ha95} P.D. Harty {\it et al.},
                Phys. Rev. C {\bf 51} (1995) 1982.
\bibitem {Ry92} J. Ryckebusch {\it et al.},
                Phys. Rev. C {\bf 46} (1992) R829.
\bibitem {Bo95} I. Bobeldijk {\it et al.},
                Phys. Lett. B {\bf 356} (1995) 13.
\bibitem {Sp90} S.V. Springham {\it et al.},
                Nucl. Phys. {\bf A517} (1990) 93.
\bibitem {Mi95} G.J. Miller {\it et al.},
                Nucl. Phys. {\bf A586} (1995) 125.
\bibitem {Ab95} C. Van den Abeele {\it et al.},
                Phys. Lett. B {\bf 296} (1992) 302.
\bibitem {Ir93} D.G. Ireland {\it et al.},
                Nucl. Phys. {\bf A554} (1993) 173.

\end{thebibliography}
\end{document}